\documentclass{appolb}
\usepackage{amssymb}
\usepackage{amsmath}%

\usepackage{graphicx}

\usepackage{hyperref}
\usepackage{lineno}

\usepackage[utf8]{inputenc}
\usepackage[normalem]{ulem}
\usepackage{color}
\usepackage{xspace}

\newcommand{\der}{\ensuremath{ {\rm d}}}

\newcommand{\pt}{\ensuremath{p_{\rm T}}\xspace}

\begin{document}



\title{Statistical hadronization of b-quarks in Pb-Pb collisions at LHC energy: a case for partial equilibration of b-quarks?
}

\author{A.~Andronic
\address{Westf\"alische Wilhelms-Universit\"at M\"unster, Institut f\"ur Kernphysik, M\"unster, Germany}
\\[3mm]
{P.~Braun-Munzinger
  \address{Research Division and ExtreMe Matter Institute EMMI, GSI Helmholtzzentrum f\"{u}r Schwerionenforschung GmbH, Darmstadt, Germany; \\
    Physikalisches Institut, Ruprecht-Karls-Universit\"{a}t Heidelberg, Heidelberg, Germany}
}
\\[3mm]
{K.~Redlich
  \address{University of Wroc\l aw, Institute of Theoretical Physics, 50-204 Wroc\l aw, Poland}
}
\\[3mm]
{J.~Stachel
  \address{Physikalisches Institut, Ruprecht-Karls-Universit\"{a}t Heidelberg, Heidelberg, Germany}
}
}

  \maketitle

  \begin{abstract}
    Predictions are presented within the framework of the statistical hadronization model for integrated yields of bottomonia in Pb-Pb collisions at the LHC. We investigate the centrality dependence of $\Upsilon$ production and provide predictions for a large set of still-unmeasured open-beauty hadrons. 
  \end{abstract}

The Statistical Hadronization Model (SHM) has proved very sucessful in describing hadron yields in central nucleus-nucleus collisions over a broad range of energies  \cite{Andronic:2017pug}.
The application of the SHM for the description of charm-quark hadrons was proposed long ago \cite{BraunMunzinger:2000px} and improved ever since, see ref. \cite{Andronic:2021erx} for the most recent account. 
In our model, both open-charm hadrons and charmonia of low/moderate \pt ($\lesssim 5$ GeV/c) are formed late during the (cross over) phase transition.
The high-\pt part of charm quarks is not fully thermalized, allowing via parton energy loss the extraction of charm transport coefficients.

The present paper addresses the question of whether on can use the same description for beauty quarks and bottomonia.
We have earlier provided predictions in the SHM framework \cite{Andronic:2006ky} and we revisit and extend those in light of current understanding and compared them to bottomonium data at the LHC.  
Thermalization for beauty quarks is implied by current data at the LHC, both concerning elliptic flow $v_2$ and the nuclear modification factor $R_{AA}$ \cite{ALICE:2020hdw,ATLAS:2020yxw,ALICE:2022tji}. These data suggest that thermalization is not complete, even for the bulk (low \pt) b-quark production.
Data on $\Upsilon$ production in Pb--Pb collisions at 5.02 TeV at midrapidity by CMS \cite{CMS:2018zza} and at forward rapidity by ALICE \cite{ALICE:2020wwx} exhibit rich features, dominated by a suppression of production in Pb--Pb compared to pp collisions. The transport \cite{Du:2017qkv,Yao:2020xzw} and hydrodynamical \cite{Krouppa:2016jcl} models are quite successful in describing these features. These models implement the so-called "sequential suppression" \cite{Digal:2001ue}, in which color screening affects differently the bottomonia state, according to their respective binding energies.
Also a parametrized approach called "comover model" \cite{Ferreiro:2018wbd} describes the data (see a comprehensive comparison in \cite{ALICE:2020wwx}).
A recent theoretical advance is brought by the heavy quark quantum dynamics treatment \cite{Islam:2020bnp} (see also reviews \cite{Rothkopf:2019ipj,Akamatsu:2020ypb}).

A state-of-the-art Lattice QCD study \cite{Bala:2021fkm} shows that the width of the lowest-lying static heavy quark-antiquark bound state in QGP grows with the temperature, but different extraction methods give rather different quantitative results.

In this paper we present predictions within the statistical hadronization model for beauty quarks (SHMb) for the LHC energy $\sqrt{s_{NN}}=5.02$ TeV. The model predicts absolute yields for open beauty hadrons as well as for bottomonia. For the latter we compare to available data, calculating $R_{AA}$ employing the measured pp data. We investigate the centrality dependence and provide predictions for the still-unmeasured open-beauty hadrons.

The essential assumption of the model is the $\mathrm{b}$-quark thermalization.
As for charm, the $\mathrm{b}$-quark balance equation in the framework of canonical thermodynamics (through the ratio of Bessel functions $I_1/I_0$) contains the fugacity $g_b$, determined by the initial (direct) $b\bar b$ production yield $N_{b\bar{b}}^{dir}$ 

\begin{equation}
    N_{b\bar{b}}^{dir}=\frac{1}{2}g_b V n_{ob}^{th}
\frac{I_1(g_b V n_{ob}^{th})}{I_0(g_b V n_{ob}^{th})}
+ g_b^2 V n_{b\bar{b}}^{th}
\end{equation}

Here, $n_{ob}^{th}$ and $n_{b\bar{b}}^{th}$ are sums over all open and  hidden beauty hadrons, respectively, and are calculated in the grand canonical approach employing the parameters derived from the fit of particle yields in central Pb--Pb collisions \cite{Andronic:2017pug} ($T=156.6\pm1.7$ MeV, $\mu_B=0.7\pm 3.8$ MeV, $V=V_{\Delta y=1}=4175\pm 380$ fm$^3$, determined at 2.76 TeV; for 5.02 TeV, $V_{\Delta y=1}=4997$ fm$^3$).
The yields of any open beauty hadron ($B$) and of bottomonia ($\Upsilon$) are calculated as follows:
\begin{equation}
N_{B} = g_b V n_{B}^{th} I_1/I_0, \quad N_{\Upsilon} = g_b^2 V n_{\Upsilon}^{th}
\end{equation}

The bottom production cross section in pp collisions and an estimate of shadowing are employed, scaled by the nuclear overlap function obtained in the Glauber approach.
We use the ALICE measurement in pp at midrapidity, $\frac{\der \sigma_{\mathrm{b\bar{b}}}}{\der y}=34.5\pm 2.4(stat)^{+4.7}_{-2.9}(tot. syst)\, \mu\mathrm{b}$ \cite{ALICE:2021mgk} and FONLL calculations \cite{Cacciari:2012ny} to extrapolate to other rapidities. 
Considering a shadowing of 30\%, independent of $y$, leading for $y$=0 ($y$=2.5-4) to $\frac{\der \sigma_{\mathrm{b\bar{b}}}}{\der y}= 24.1\pm 4.8 \, (11.7\pm 2.3)\, \mu$b (a 20\% overall uncertainty is assumed).
The resulting rapidity densities of $\mathrm{b\bar{b}}$ pairs and the beauty fugacities for central collisions in two rapidity ranges are shown for illustration in Table~\ref{tab:param}. We note that the beauty fugacity values are 7 orders of magnitude larger compared to those for charm ($\simeq$30).

\begin{table}[htb]
  \begin{tabular}{l|c|c}
    ~ & $|y|<0.5$ & $y$=2.5-4 \\ \hline
    $\frac{\der \sigma_{\mathrm{b\bar{b}}}}{\der y}$ ($\mu$b) & $24.1\pm 4.8$ & $11.7\pm 2.3$ \\
    $\frac{\der N_{\mathrm{b\bar{b}}}}{\der y}$ (0-10\%) & 0.57 & 0.28 \\
    $g_b$ (0-10\%) & $1.05\cdot10^9$ & $0.86\cdot10^9$\\
    \end{tabular}
\caption{SHMb parameters for midrapidity and forward rapidity for 0-10\% Pb-Pb collisions at 5.02 TeV.} \label{tab:param}
\end{table}

The model includes a corona component, calculated in the Glauber approach as a function of centrality; we assume corona in case the mean nuclear density of any of the 2 colliding nuclei is $\rho<0.1\rho_0$. Here, production as in pp collisions is assumed. We employ measured $\Upsilon$ data and, for the rapidity dependence, arbitrarily-scaled FONLL calculations for $\frac{\der \sigma_{\mathrm{b\bar{b}}}}{\der y}$.
The $\Upsilon$ data in pp are used as well for the calculation of the the nuclear modification factor $R_{AA}$ based on the model prediction of absolute yields in Pb-Pb (as a function of centrality).

The beauty-hadron spectrum employed in our calculations includes all states listed in the PDG \cite{ParticleDataGroup:2020ssz}, comprising 48 individual meson states and 46 baryon states. 
The presence of currently-unknown open-beauty states will lead in our model to a reduction of the predicted bottomonia yields.

\begin{figure*}[h!]
 \begin{tabular}{lr} \begin{minipage}{.49\textwidth}
\centerline{\includegraphics[width=.95\textwidth]{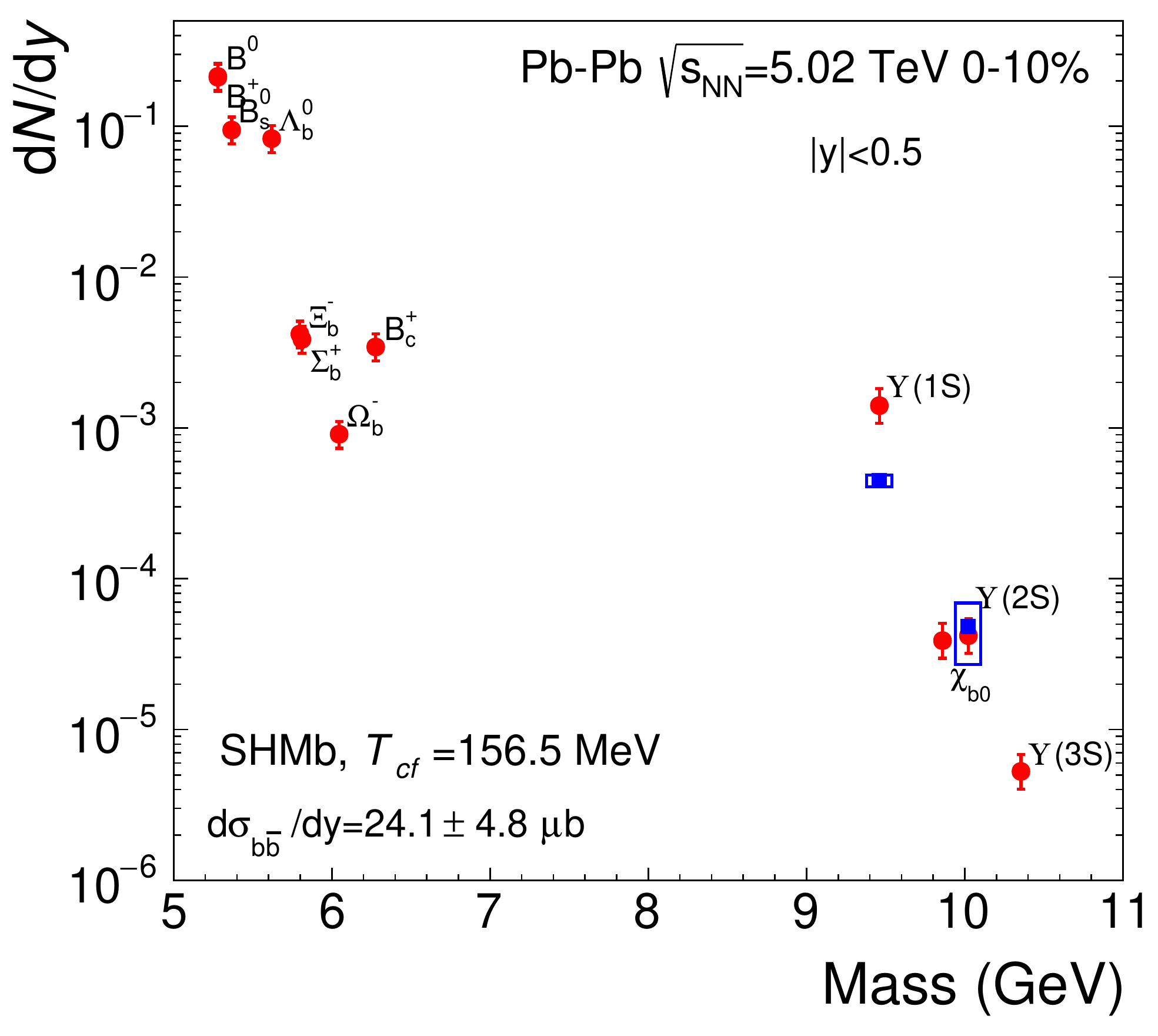}}
   \end{minipage} & \begin{minipage}{.49\textwidth}
\centerline{\includegraphics[width=.95\textwidth]{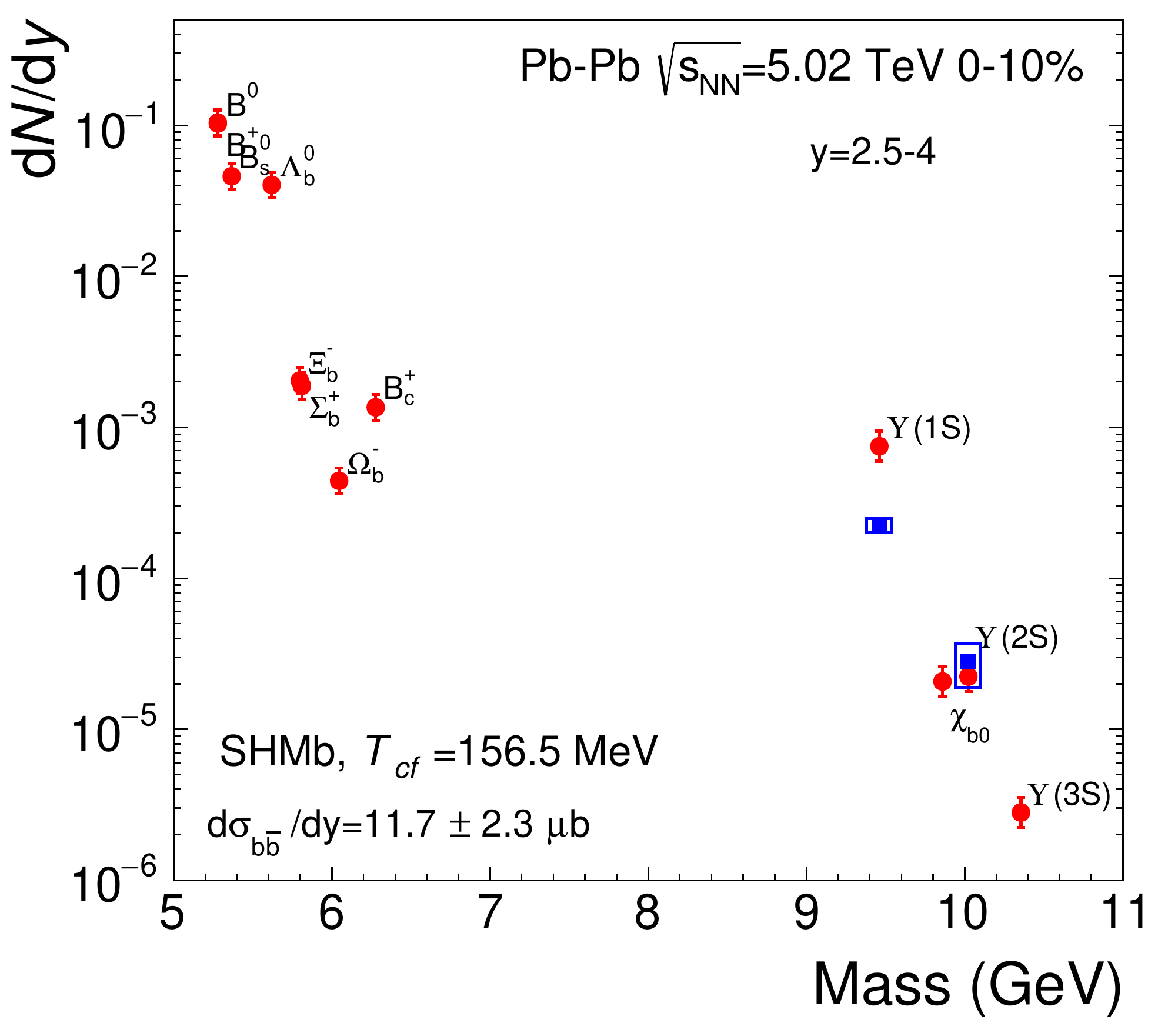}}
     \end{minipage}\end{tabular}
  \caption{Yields (rapidity density) for open beauty and bottomonia at midrapidity (left) and forward rapidity (right) for central collisions.} \label{fig:dndy}
\end{figure*}

The yields for various open beauty and bottomonia species are shown for central collisions in Fig.~\ref{fig:dndy}, for midrapidity and forward rapidity.
The power of the model to predict the beauty sector is evident.
The $\Upsilon(1S)$, $\Upsilon(2S)$ data are included. In lack of published absolute yields, we calculated them employing the published $R_{AA}$ values \cite{CMS:2018zza,ALICE:2020wwx} and the pp reference.
The model describes the $\Upsilon(2S)$ data quite well, however, overestimates the $\Upsilon(1S)$ data significantly.

\begin{figure*}[h!]
 \begin{tabular}{lr} \begin{minipage}{.49\textwidth}
\centerline{\includegraphics[width=.95\textwidth,height=.7\textwidth]{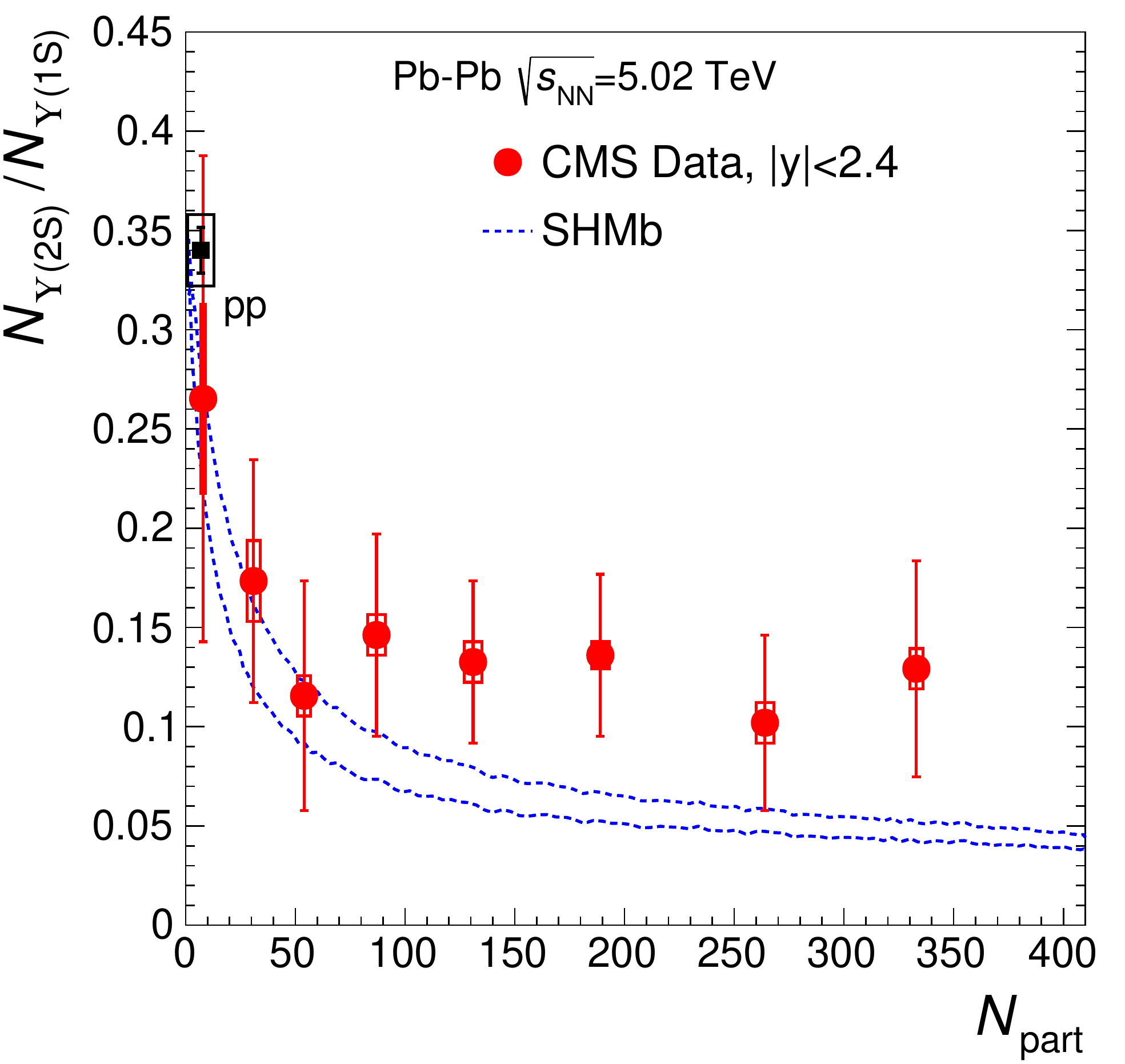}}
   \end{minipage} & \begin{minipage}{.49\textwidth}
\centerline{\includegraphics[width=.95\textwidth,height=.7\textwidth]{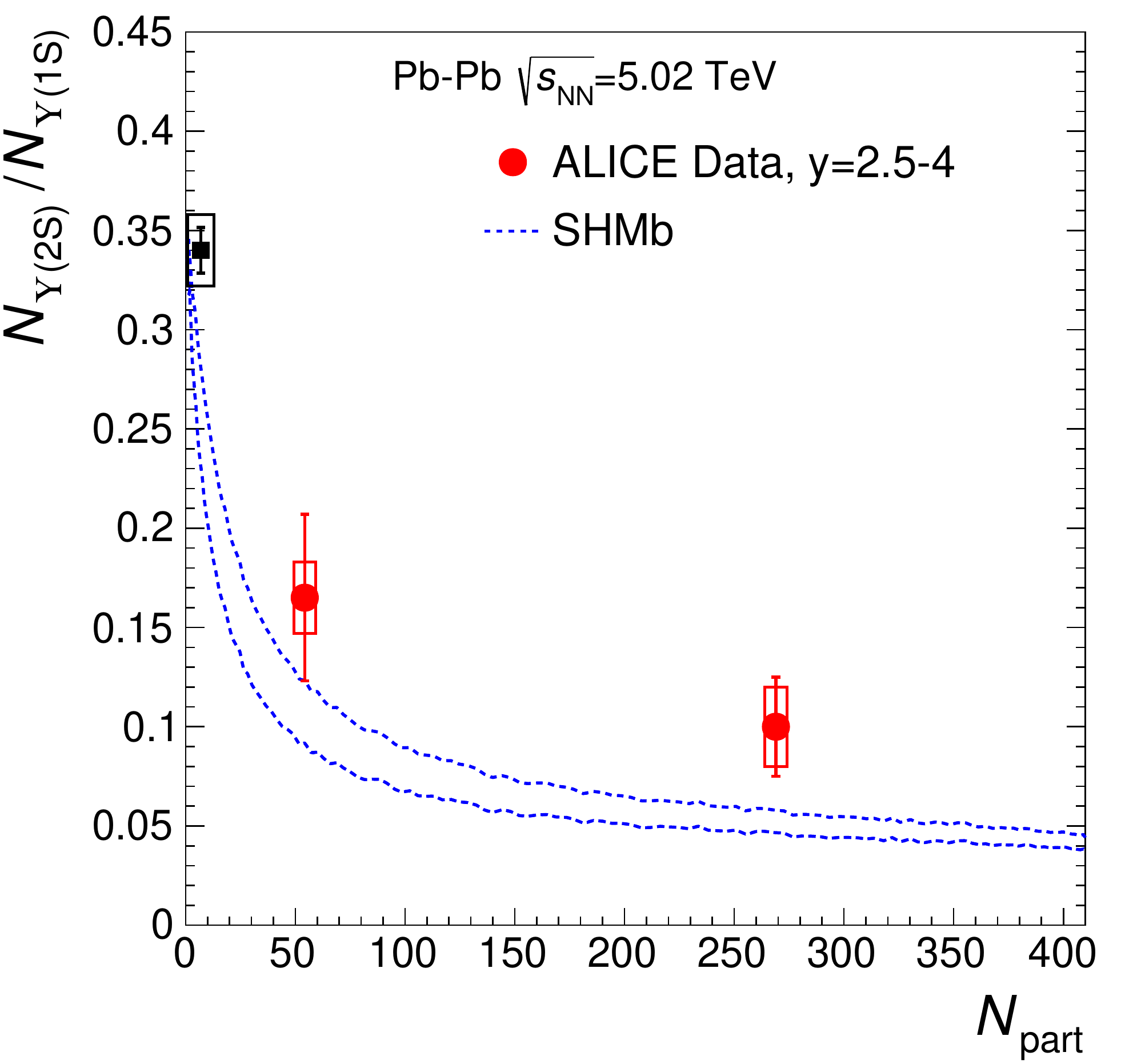}}
     \end{minipage}\end{tabular}
 \caption{Ratio of yields of $\Upsilon(2S)$ and $\Upsilon(1S)$ mesons for midrapidity (left) and forward rapidity (right). Here the upper curve is exemplarily for a larger fraction of corona, $0.2\rho_0$.} 
 \label{fig:yratio}
\end{figure*}

The ratio of yields of $\Upsilon(2S)$ and $\Upsilon(1S)$ mesons is compared to data at midrapidity and forward rapidity in Fig.~\ref{fig:yratio}. As expected from the comparison of the absolute yields, the model underpredicts the ratio.

\begin{figure*}[h]
 \begin{tabular}{lr} 
 \begin{minipage}{.49\textwidth}
\centerline{\includegraphics[width=.95\textwidth,height=.7\textwidth]{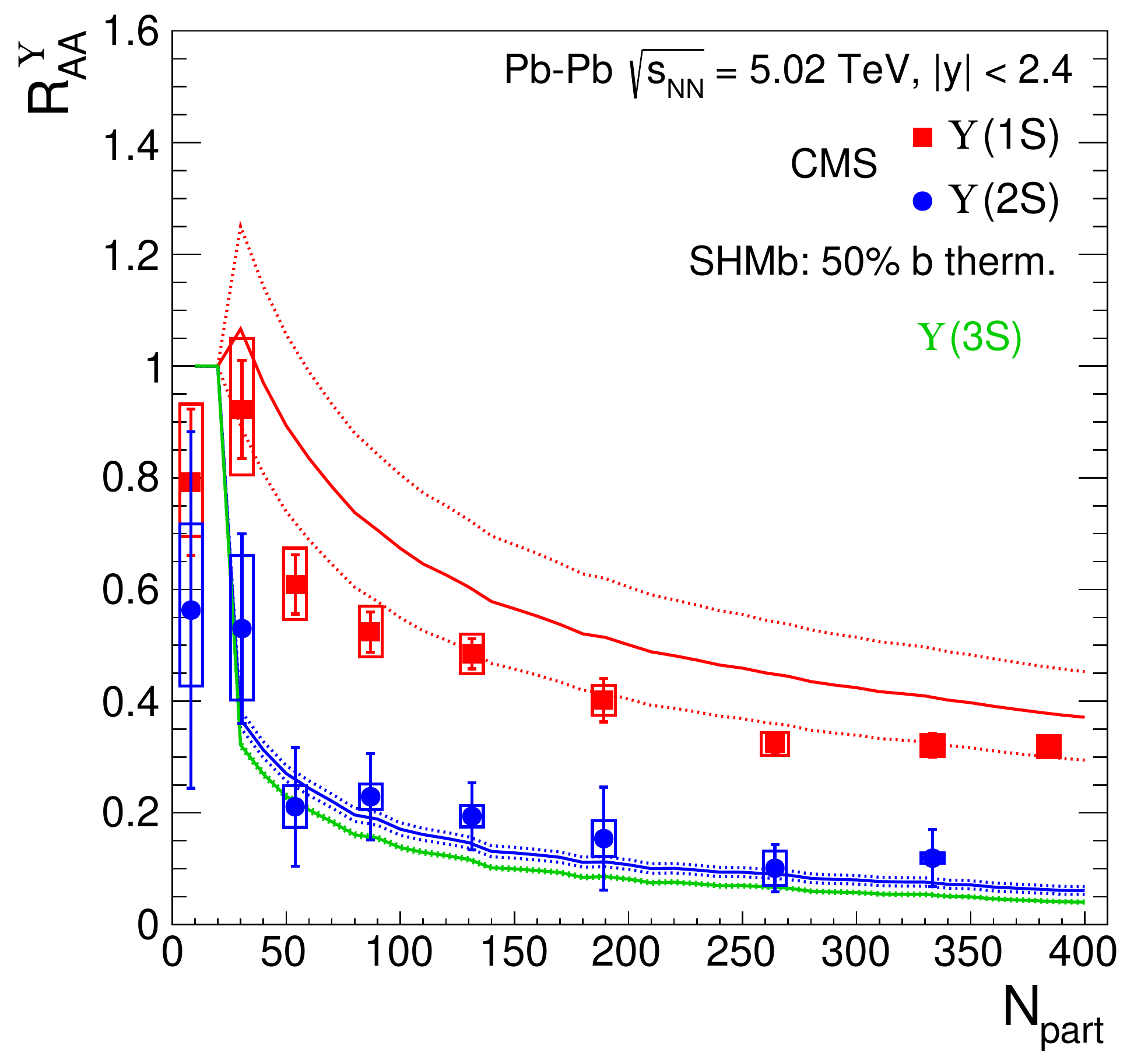}}
   \end{minipage} & \begin{minipage}{.49\textwidth}
\centerline{\includegraphics[width=.95\textwidth,height=.7\textwidth]{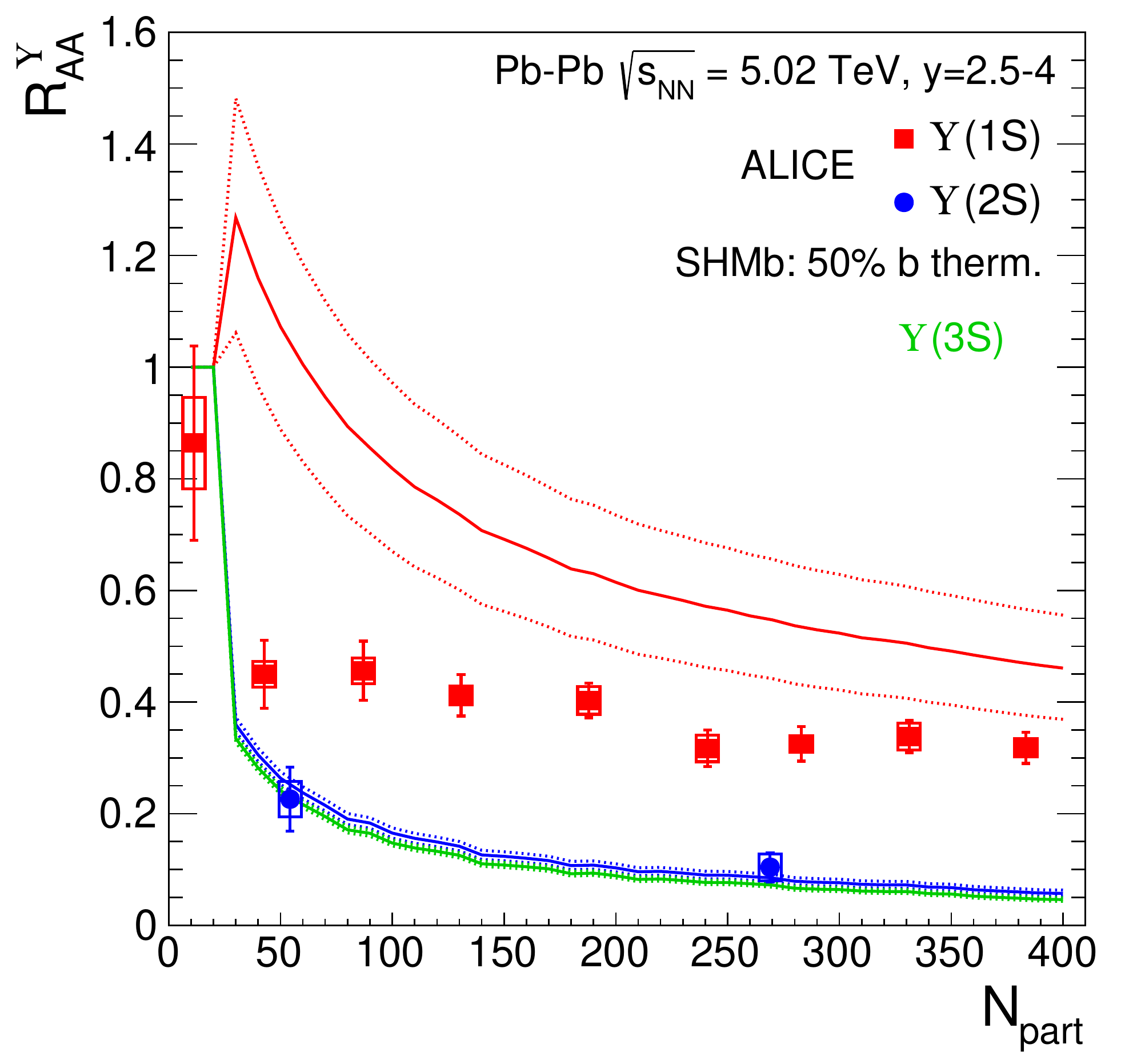}}
     \end{minipage} \\
\begin{minipage}{.49\textwidth}
\centerline{\includegraphics[width=.95\textwidth,height=.7\textwidth]{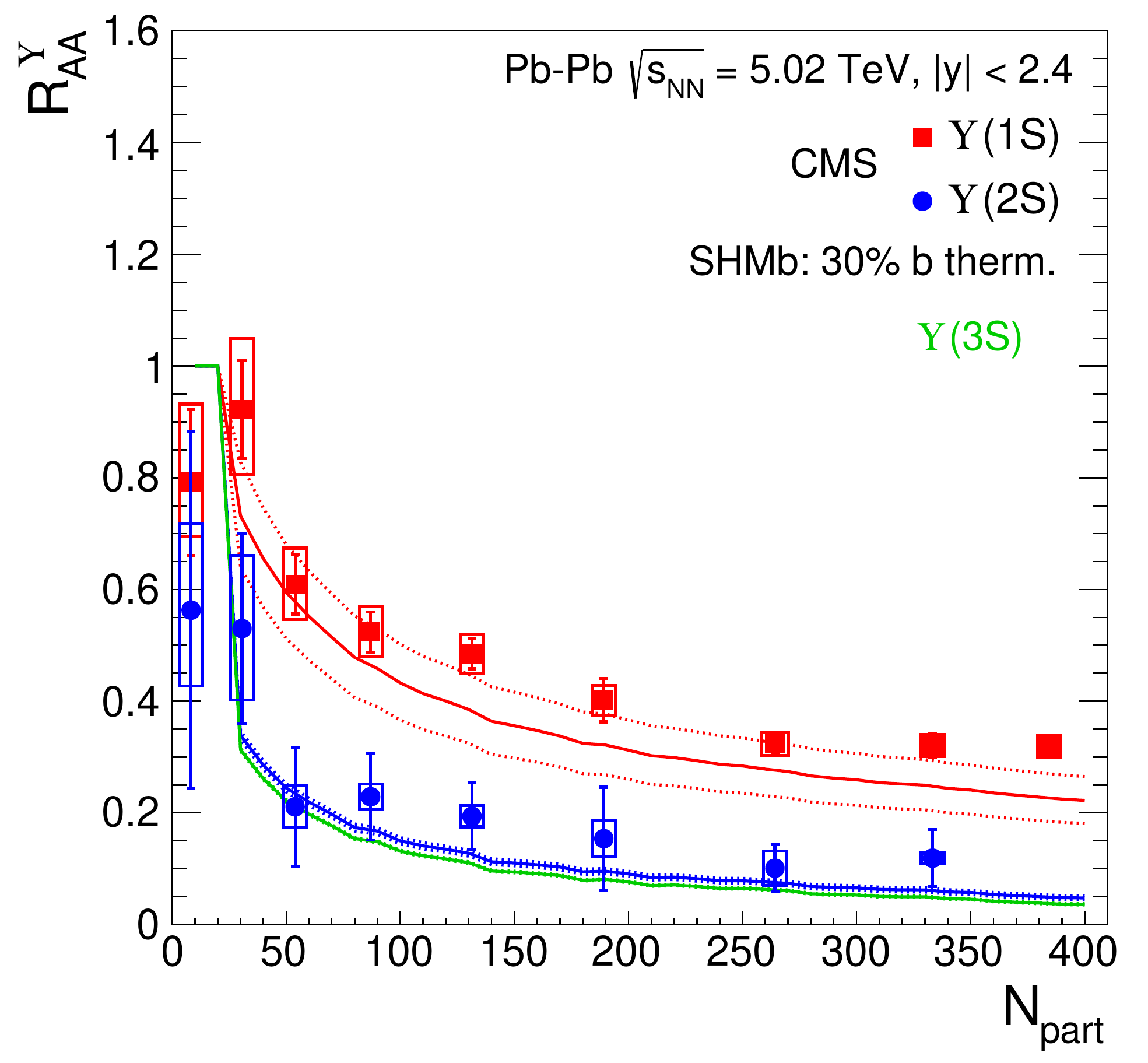}}
   \end{minipage} & \begin{minipage}{.49\textwidth}
\centerline{\includegraphics[width=.95\textwidth,height=.7\textwidth]{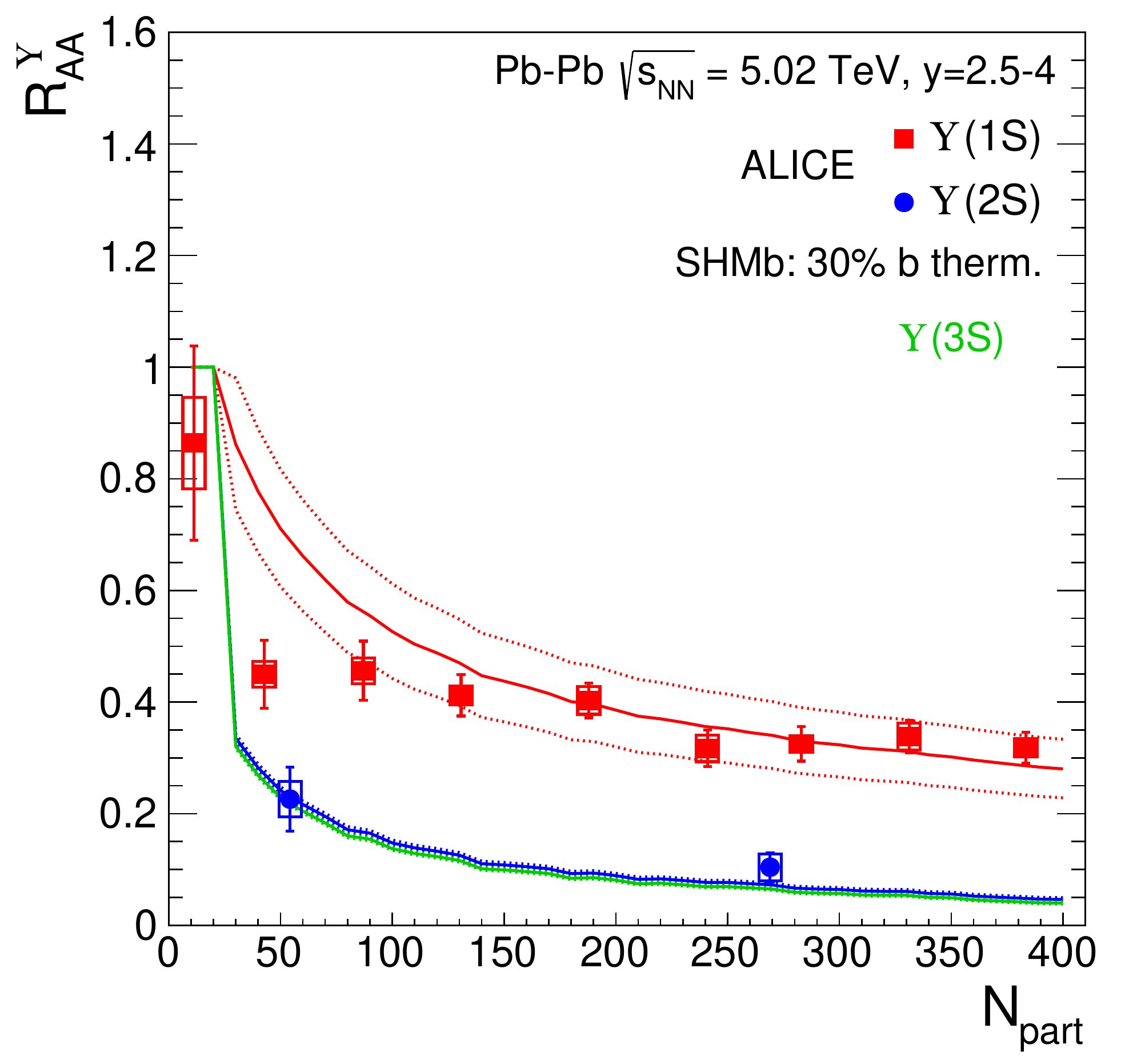}}
     \end{minipage}
 \end{tabular}
  \caption{The nuclear modification factor of $\Upsilon$ mesons for midrapidity (left column) and forward rapidity (right column). The upper row is for 50\% beauty thermalization , the lower row is for 30\%} \label{fig:raa}
\end{figure*}

To explore the effect of a partial thermalization of b quarks, we introduce in the model an arbitrary suppression of beauty pairs available for statistical hadronization at the phase boundary. 
The cases of 50\% and 70\% suppression are presented in Fig.~\ref{fig:raa}, where we compare for the $\Upsilon$ mesons our predictions and the measured data.
The data are well described for the case of 30\% beauty thermalization.

Obviously, the question arises why the non-thermalized beauty component would not bring any contribution to the $\Upsilon$ yield.
One can argue that the strongly coupled QGP leads to a decorrelation of the $\mathrm{b}$ and $\mathrm{\overline{b}}$ quarks. But for regeneration of $\Upsilon$ mesons from these $\mathrm{b}$ quarks, full local thermalization is required, a much stronger condition than decorrelation.

It is important to note that the presence of currently-unknown open-beauty states will lead in our model to a reduction of the predicted bottomonia yields. Consequently, the estimated 30\% thermalization fraction deduced here is expected to increase in our model if new open-beauty states will be included in the future.

We conclude that a full thermalization of beauty quarks in the QGP does not explain the bottomonium data at the LHC. A thermalization fraction of about 30\% does, but it is important to note that the presence of currently-unknown open-beauty states will lead in our model to a reduction of the predicted bottomonia yields.

\vspace{3mm}
\textbf{Acknowledgment}
This   work   is   part   of   and   supported   by   the DFG   Collaborative   Research   Centre   “SFB   1225 (ISOQUANT)”.
K.R. acknowledges the support by the Polish National Science Center under the Opus grant no.  2018/31/B/ST2/01663, and the Polish Ministry of Science and Higher Education.

\bibliography{qm22} 

\end{document}